\documentclass[conference]{IEEEtran}
\IEEEoverridecommandlockouts
\usepackage{cite}
\usepackage{amsmath,amssymb,amsfonts}
\usepackage{algorithmic}
\usepackage{graphicx}
\usepackage{textcomp}
\usepackage{xcolor}
\usepackage{tabularx}
\usepackage{caption}
\usepackage{subcaption}
\usepackage{cleveref}

\title{Bibliography management: \texttt{natbib} package}
\def\BibTeX{{\rm B\kern-.05em{\sc i\kern-.025em b}\kern-.08em
    T\kern-.1667em\lower.7ex\hbox{E}\kern-.125emX}}

\makeatletter
\newcommand{\linebreakand}{%
  \end{@IEEEauthorhalign}
  \hfill\mbox{}\par
  \mbox{}\hfill\begin{@IEEEauthorhalign}
}
\makeatother

\def\BibTeX{{\rm B\kern-.05em{\sc i\kern-.025em b}\kern-.08em
    T\kern-.1667em\lower.7ex\hbox{E}\kern-.125emX}}
\begin{document}
\title{Designing a Visual Cryptography Curriculum for K-12 Education\\
}
\author{\IEEEauthorblockN{Pranathi Rayavaram\IEEEauthorrefmark{1}, Sreekriti Sista\IEEEauthorrefmark{2}, Ashwin Jagadeesha\IEEEauthorrefmark{1}, Justin Marwad\IEEEauthorrefmark{1},
Nathan Percival\IEEEauthorrefmark{1}, \linebreakand Sashank Narain\IEEEauthorrefmark{1}, and Claire Seungeun Lee\IEEEauthorrefmark{1}}
\IEEEauthorblockA{\IEEEauthorrefmark{2}Heritage Highschool, Frisco, TX, USA\\}
\IEEEauthorblockA{\IEEEauthorrefmark{1}University of Massachusetts Lowell, Lowell, MA, USA\\
}
nagapranathi\_rayavaram@student.uml.edu, sreekriti.sista.912@k12.friscoisd.org, 
ashwin\_jagadeesha@student.uml.edu, \\ justin\_marwad@student.uml.edu, nathan\_percival@student.uml.edu,
sashank\_narain@uml.edu, and claire\_lee@uml.edu
}

\maketitle
\IEEEpubidadjcol

\begin{abstract}
We have designed and developed a simple, visual, and narrative K-12 cybersecurity curriculum leveraging the Scratch programming platform to demonstrate and teach fundamental cybersecurity concepts such as confidentiality, integrity protection, and authentication. The visual curriculum simulates a real-world scenario of a user and a bank performing a bank transaction and an adversary attempting to attack the transaction. We have designed six visual scenarios, the curriculum first introduces students to three visual scenarios demonstrating attacks that exist when systems do not integrate concepts such as confidentiality, integrity protection, and authentication. Then, it introduces them to three visual scenarios that build on the attacks to demonstrate and teach how these fundamental concepts can be used to defend against them. We conducted an evaluation of our curriculum through a study with 18 middle and high school students. To evaluate the student's comprehension of these concepts we distributed a technical survey, where overall average of students answering these questions related to the demonstrated concepts is 9.28 out of 10. Furthermore, the survey results revealed that 66.7\% found the system extremely easy and the remaining 27.8\% found it easy to use and understand.

\end{abstract}


%(appr. 20\%) Briefly describe the motivation behind your study or change in practice (that is, what you wanted to find out).  This section should include a brief discussion that identifies your research question, hypothesis or goal of the study, or the reason behind the change in practice described in this paper.

\section{\textbf{Introduction}}
Our data, digital devices, resources, and national secrets are more susceptible to cybercrime than ever before. Even then, there is a cybersecurity workforce shortage and a general lack of awareness that is required to defend against these crimes/attacks. The International Information System Security Certification Consortium $(ISC)^2$ estimates that the cybersecurity workforce must increase by 145\% to fulfill global demand ~\cite{isc2_cybersecurity_2021}. Currently, there are around 4 million vacant positions due to the talent gap. Bridging this cybersecurity talent gap is essential and has become one of the top priorities globally~\cite{cybersecurity_summit_report}. To address this problem, many government organizations and cybersecurity experts are ensuring to bring cybersecurity curriculum to the K-12 education level. These initiatives aim to encourage the next generation to pursue cybersecurity as a career and build the required skill set at an early age to enter into the cyber workforce. In addition, this also improves cybersecurity literacy nationwide and increases awareness among children. This is important as cybercrime is increasingly more prevalent than before, and it is, thus, crucial that they know and understand how to secure their private data.

%At the same time, we believe that the COVID-19 pandemic has accelerated the adoption of digital technologies among young children. The sudden transition to virtual platforms has caused an unintended effect that K-12 students now spend a much longer time in front of their screens. Oftentimes, such interactions with the digital world are security-sensitive in nature. Some examples include but are not limited to cyberbullying, falling prey to cyber predators, posting private information (PII) online, and phishing. As a case in point, a recent national survey of 918 K-12 educators revealed that cybersecurity education is underrepresented in the K-12 curriculum. In particular, cryptography is even scarcer, with only 5\% of high school students and less than 1\% of middle-school students reporting having knowledge on the topic~\cite{state_art}. We believe that this lack of cybersecurity awareness can pose significant threats and severe consequences to young children. The broader integration of cybersecurity education into the K-12 curriculum is widely believed to be an appropriate solution to this problem.

For the development of a K-12 cybersecurity curriculum, organizations such as cyber.org have collaborated with K-12 educators, the cybersecurity industry, and the government to create the very first voluntary national cybersecurity standards ~\cite{cyber.org}. These standards provide curriculum developers and teachers at schools with clear paths and resources they need in order to provide robust cyber education for every student. In addition, a core theme for security has been included in this curriculum, which comprises of sub-concepts such as the Confidentiality, Integrity, and Accessibility (CIA) triad, Access Control (ACC), Data Security (DATA), Threats and Vulnerabilities (INFO), and Cryptography (CRYP). Unfortunately, while these concepts are important to cybersecurity curricula, K-12 educators end up using complex tools or programming languages such as Python to teach these concepts to K-12 students due to a relative lack of age-appropriate, intuitive, and easy-to-use tools. Furthermore, according to a nationally representative survey from cyber.org administered by the EdWeek Research Center ~\cite{cyberorg_state_2020}, 80\% of rural educators claim that their students lack adequate cybersecurity resources. This includes a lack of access to cybersecurity education, from coursework to educators knowledgeable about the topic. These communities, which are disproportionately rural and low-income, are a crucial starting point for improving access to cybersecurity education in order to provide equal opportunities for students choosing a cybersecurity career.

\begin{figure}[t]
     \begin{subfigure}[b]{0.43\textwidth}
        \centering
        \includegraphics[width=0.88\linewidth]{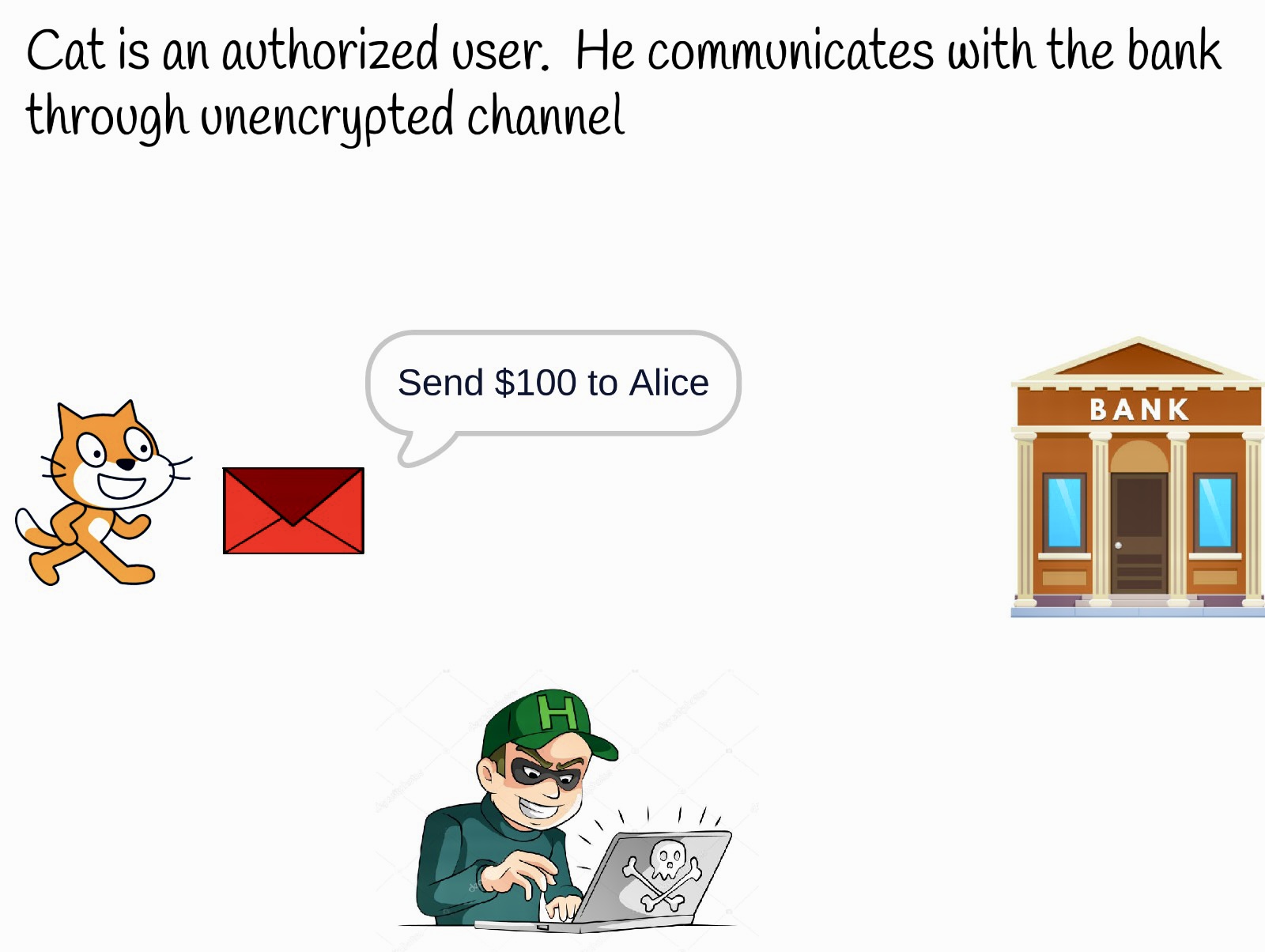}
        \caption{User cat sends a transaction message to a bank.}
        \label{fig:scenario1a}
      \end{subfigure}
     \hfill
     \newline
     \newline
     \begin{subfigure}[b]{0.43\textwidth}
        \centering
        \includegraphics[width=0.88\linewidth]{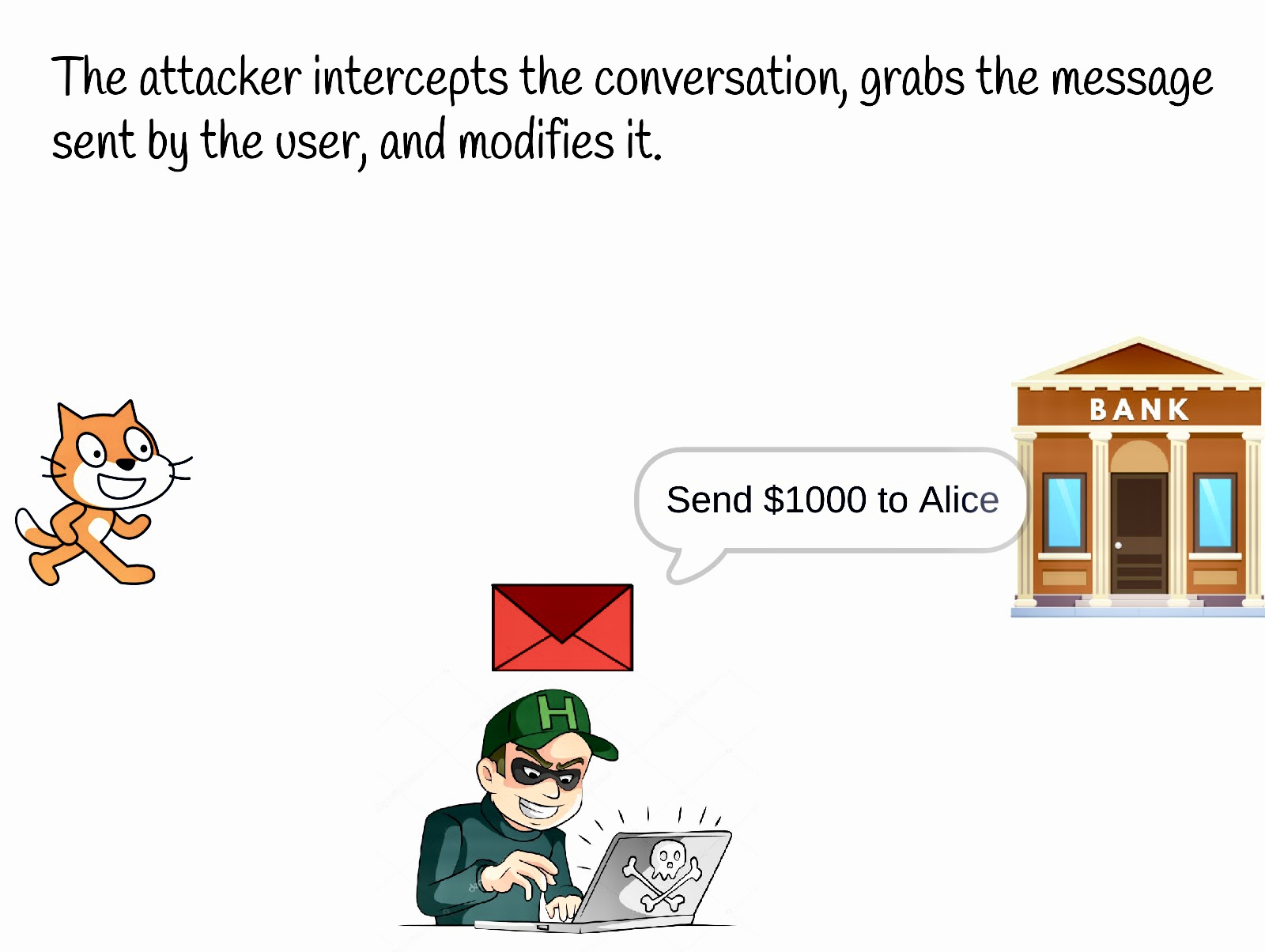}
        \caption{An attacker modifies the transaction message.}
        \label{fig:scenario1b}
      \end{subfigure}
     \hfill
    \newline
    \newline
    \begin{subfigure}[b]{0.43\textwidth}
        \centering
        \includegraphics[width=0.88\linewidth]{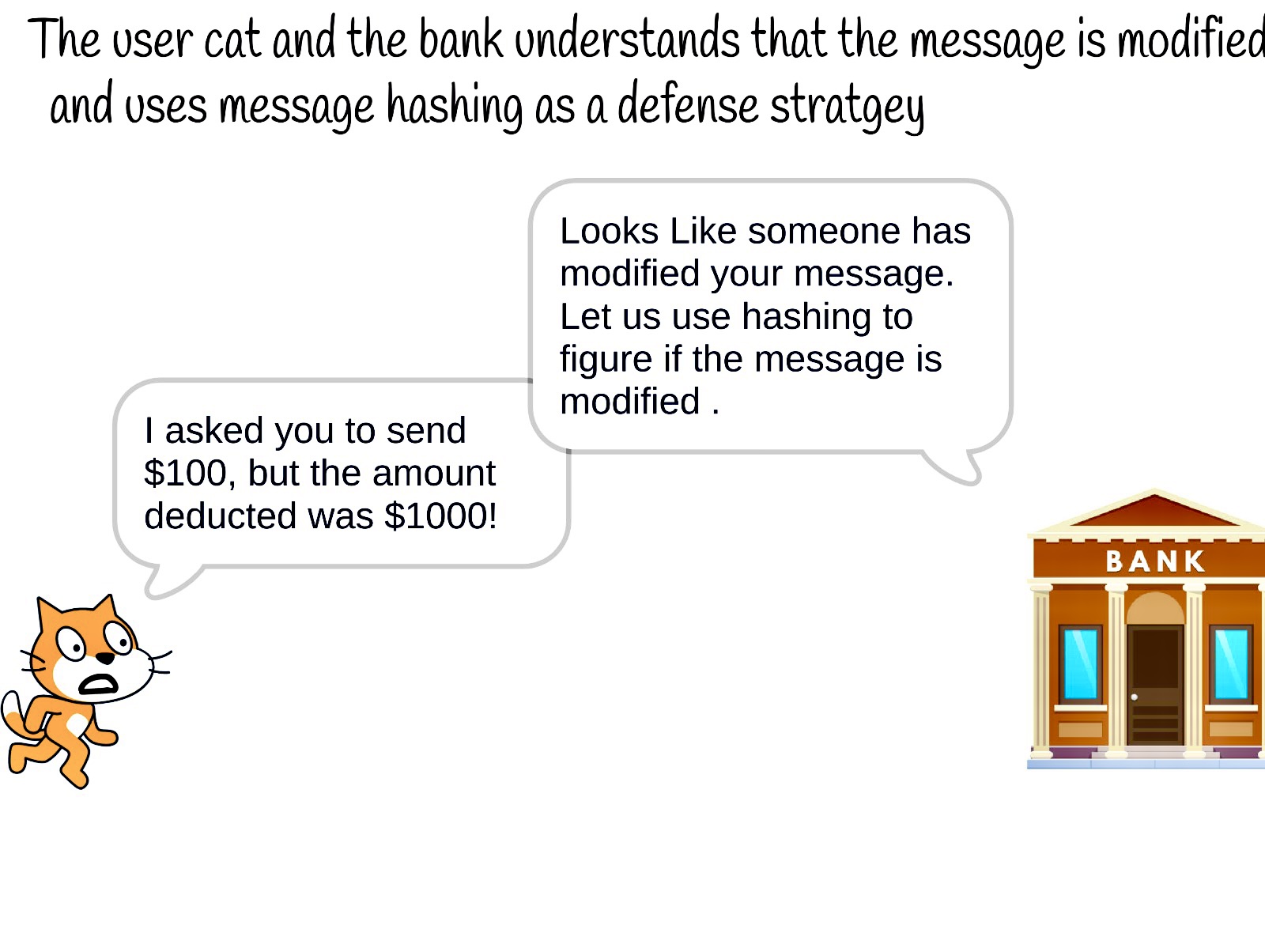}
        \caption{The user and bank discuss message integrity.}
        \label{fig:scenario1b}
    \end{subfigure}
      
    \caption{A scenario of our visual curriculum where an attacker changes a user’s transaction from \$100 to \$1000. This message integrity breach forces the user and bank to think about message hashing as a protection against tampering.}
    \label{fig:Scenario1}
\end{figure}

To address the aforementioned problems, we have developed a simple, visual, and intuitive cybersecurity curriculum on Scratch for K-12 students. Scratch is a financially feasible and globally accessible platform, which implies that developing the curriculum on Scratch would provide equitable educational chances for students from low-income or rural regions. Scratch's block-based programming and graphical representations encourage mathematics and science education. For instance, Harvard has integrated a Computer Science course using the Scratch platform ~\cite{harvard} to teach programming fundamentals. Given these considerations, we believe that presenting a curriculum on Scratch would be appropriate for both teachers and students.

Our research focuses on answering the following research questions (RQs). First, \textbf{\textit{RQ1: Can we develop a cybersecurity curriculum using simple technologies that enables K-12 children to intuitively understand and apply core cybersecurity concepts such as confidentiality, integrity protection, and authentication?}} Second, \textbf{\textit{RQ2: Can we design and implement this curriculum in a manner where students can understand these concepts under minimal or no supervision?}}

In response to RQ1 and RQ2, we used Scratch - the block-based coding platform - to design a novel cybersecurity curriculum focused on the basics of cybersecurity, such as confidentiality, integrity protection, and authentication. According to the K-12 national cybersecurity curriculum from cyber.org ~\cite{cybersecurity_summit_report}, these are fundamental concepts that must be taught to everyone beginning cybersecurity education. We believe that understanding these concepts leads to a stronger understanding of modern cybersecurity schemes such as TLS and SSH, the technologies that protect our day-to-day online activities.
Using the visual capabilities of Scratch, we created six scenarios that illustrate the story of an attacker attacking a system that a user and a bank are using to perform a banking transaction. A user connects with the bank to make a transaction with his friend, and an attacker is trying to attack (read, modify, and/or fabricate) the transaction the user has made. Each time the attacker comes up with a new strategy to attack the transaction, the user and bank defend against those attacks by introducing the concepts of confidentiality, integrity protection, and authentication. For instance, Fig.~\ref{fig:Scenario1} shows how the concept of integrity protection can be demonstrated to young children simply by building a visual story-like environment in Scratch. In this example, a user (cat) and a bank perform a banking transaction in which “Send \$100 to Alice” gets modified by an attacker to  “Send  \$1000  to  Alice”. The user and bank think of message hashing as a protection mechanism against such modification. We believe that this real-time visual narrative will assist students in comprehending the significance and existence of these fundamental system security concepts. The capability of our curriculum to clearly show offensive and defensive scenarios will assist students in comprehending and excelling in this field. We anticipate that in the future, if children are educated with more sophisticated cybersecurity concepts, they will be able to create these real-time secure infrastructure scenarios on the Scratch platform. This enables students to comprehend the intuitive application of cybersecurity concepts in real-world without any programming or tool-specific complexities.

 To assess the effectiveness of our curriculum, we recruited 18 middle and high school students (10 girls and 8 boys) for a 3-hour workshop. During this time, the students were taught the previously mentioned core cybersecurity concepts using our curriculum. Afterwards, 
 %The first hour of the workshop was dedicated to teaching the basics of cybersecurity, while the remaining time was spent presenting the concepts of confidentiality, integrity, authentication, and anti-replay through our.
 we administered a survey questionnaire in order to assess the students' comprehension of the topics that were taught in the workshop. We also included secure and insecure visual scenario questions about these concepts in the survey. 
 %These questions prompt the user/kids to think which scenario is secure and why the other is insecure. 
 In this three-hour session, the students could easily understand the concepts of confidentiality, integrity protection, and authentication, and provide accurate answers to various technical questions on these concepts, with an average score of 9.28 out of 10. According to the post-survey results, 66.7\% of students said the curriculum was really easy, and 27.8\% of students considered it to be easy. The curriculum was well received by the participants, considering they would recommend it to others. Specifically, 61.1\% of students were inclined to strongly recommend this intuitive curriculum to their peers, while the remaining students were highly likely to suggest it. Our results also indicate that this style of narrative will aid students in comprehending cybersecurity concepts with minimum supervision, which might be especially helpful for those who lack access to educational resources. Our research contributions may have far-reaching effects on K-12 education by raising students' awareness about cybersecurity and helping educators fill a critical resource vacuum in their cybersecurity instruction.

\vspace{0.25em}\noindent
In summary, we make the following contributions: 
\begin{itemize}
 
\item We designed and implemented a simple, visual, and intuitive cybersecurity curriculum for K-12 students on the Scratch platform. Our initial efforts are focused on teaching children fundamental cybersecurity concepts such as confidentiality, integrity, and authentication. 
%\item We have designed a simple, visual and intuitive cybersecurity curriculum on the Scratch platform focused on teaching the core concepts of confidentiality, integrity protection, and authentication to K-12 students.

%We have designed 8 scenarios as a real-world story demonstrating the attacks and introducing the cybersecurity principles Confidentiality, Integrity, Authentication and Anti-replay concepts as defensive mechanisms to those attacks.
%We have implemented a real time situation on Scratch interface, by making use of Scratch visual capabilities and cybersecurity algorithm blocks. The situation is between a user, bank, and attacker attempting to intercept discussions. We have created 8 scenarios demonstrating the attacks and introducing the cybersecurity principles Confidentiality, integrity, Authentication and Anti-replay concepts as defensive mechanisms to those attacks.

\item We conducted a user study with 18 middle and high school students to teach them cybersecurity concepts using our curriculum. Our results show that our visual curriculum greatly simplifies the understanding of core cybersecurity concepts and is engaging to K-12 students. 
%to evaluate their comprehension of concepts as well as their perceptions of the interface.

\end{itemize}
The rest of this paper is structured as follows. In Section II, we discuss some of the related works that focus on cybersecurity curricula and tools that teach cybersecurity concepts. Section III focuses on the design and implementation of our attack and defense scenarios. Section IV describes the design, metrics, and results of the user study that was conducted with 18 middle to high school students to evaluate the effectiveness of our curriculum. We highlight the potential benefits and contributions of this current study in Section V. We then conclude our work in Section VI.

\section{\textbf{Related Work}}
We acknowledge that many initiatives exist that focus on cybersecurity education and awareness for diverse socioeconomic groups, education of women, and underrepresented groups \cite{longitudanal} \cite{securing}. This section highlights only a subset of these initiatives directly relevant to our curriculum. Our focus is on demonstrating how these initiatives encourage cybersecurity education and on tools that teach these concepts.

Recent studies have focused substantially on incorporating cybersecurity education into K–12 curricula for children \cite{exploring}. Most of the research focuses on the barriers preventing educators from delivering effective cybersecurity education in formal educational settings. All of these research projects were conducted in an effort to make cybersecurity more accessible and intuitive for young students. In order to provide a comprehensive cybersecurity curriculum for K-12 students, it is crucial to figure out how to teach them and devise an effective method of teaching. 
%A substantial number of studies have been conducted on this topic.

%The most effective way to teach students cybersecurity is by having a qualified, well-experienced, and effective teacher. The importance of teacher education programs~\cite{teacher_2011}, in this regard, comes from the fact that they train teachers in advanced cybersecurity principles and other strategies for developing robust integrated curricula. 

In the past few years, important initiatives and research have existed from the public sector, private sector,  universities, and beyond. 
As a good example that how a university can engage with K-12 students in cybersecurity, Stony Brook University's cyberMiSTS  project team designed a professional development course for middle school teachers that incorporated a recent understanding of cybersecurity based on research~\cite{cybermists}. Hacker High school from the Institute for Security and Open Methodologies (ISECOM)~\cite{isecom_hacker_2021} provides a self-directed learning curriculum focused on attack and defense skills. The public sector also plays an important role in giving young students an opportunity to be more educated and trained in cybersecurity. The AFA CyberPatriot program by the Air Force Association~\cite{cyberpatriot} and NSF GenCyber ~\cite{gencyber_gencyber_2020} are such programs that work to inspire students to pursue careers in cybersecurity or related STEM fields. This work~\cite{storytelling} focuses on teaching students about cybersecurity and developing their cognition through storytelling using social robots. Another work's ~\cite{casestudy} purpose is to include case studies in cybersecurity training. These case studies are actual breaches that have occurred in the real world, and these attacks were taught from beginning to end. In addition, efforts have been made to educate women about cybersecurity, and research suggests that applications inherent to cybersecurity will likely increase their interest in the field ~\cite{codebreakers_2020,girlsgocyberstartwebsite,womenscyberjutsu_2020,cybersecurity_minorities,Bulldog,supersleuth}.

There has been a significant research recently in developing innovative curricula and pedagogical tools in the area of cybersecurity education ~\cite{literature_review}. These tools increase the scope of education and resources to teach. These tools have developed a special interest in the research community, and their objectives generally involve creating challenges and to bring hands-on experience around specific concepts~\cite{networkrobots}. For example,~\cite{cybersec_curiculum} has focused on creating cybersecurity curriculum challenges that include defensive measures such as encryption, secure key exchange, and sequence numbers to prevent cyber attacks during robot operations. Video games for cybersecurity are also considered one of the effective techniques to provide cybersecurity training to children~\cite{Game_1,Game_2,Game_4}. Some other  works~\cite{cybervr,cybar}  focus on creating VR and AR technologies to effectively increase cybersecurity awareness. Other works such as~\cite{publickey_2010} attempt to develop visual techniques to teach public key infrastructure concepts. The Bits N Bytes organization~\cite{bits_n_bytes_bits_2020} is dedicated to bringing awareness of the field of cybersecurity to young students. Works such as Cybersecurity Lab as a Service (CLaaS)~\cite{claas_2015} have also looked at creating services that can run other cybersecurity tools on their platform.

Some of the initiatives focused on teaching cryptography through an online website, the presentation of workshops, and the distribution of cryptography challenges. Practical Cryptography~\cite{practical_cryptography} is an online lesson-based website that emphasizes the practical application of classical cryptography. In a similar line, The CryptoClub Project \cite{beissinger_about_2021} provides educational resources for use in the classroom as well as an interactive website to educate students on typical cryptographic techniques. A workshop held by Lincoln Laboratory \cite{mit_lincoln_laboratory_llcipher_2021} has the intention of teaching core cryptography theory to students who are interested in the mathematics aspects of cybersecurity. In a similar manner, NCC Group hosts Cryptopals Crypto Challenges \cite{northrop_grumman_foundation_what_2013} involving the participants in programming to address cryptography challenges. The intended audience for Cryptopals is comprised of experienced programmers interested in independently studying cryptographic concepts.

Despite the existing and ongoing efforts, our work differs from the above curricula and tools as we focus on a self-expressive, visual, and intuitive cybersecurity curriculum using the popular Scratch platform. We focus on building visual and intuitive stories for young students that aid their understanding of the fundamentals that drive cybersecurity and cryptosystems. We note that our curriculum can be integrated into the above initiatives to teach the concepts of confidentiality, integrity protection, and authentication to K-12 students.

\section{\textbf{Design of the curriculum}}\label{sec:technical}

\begin{figure}[t]
     \begin{subfigure}[b]{0.43\textwidth}
        \centering
        \includegraphics[width=0.85\linewidth]{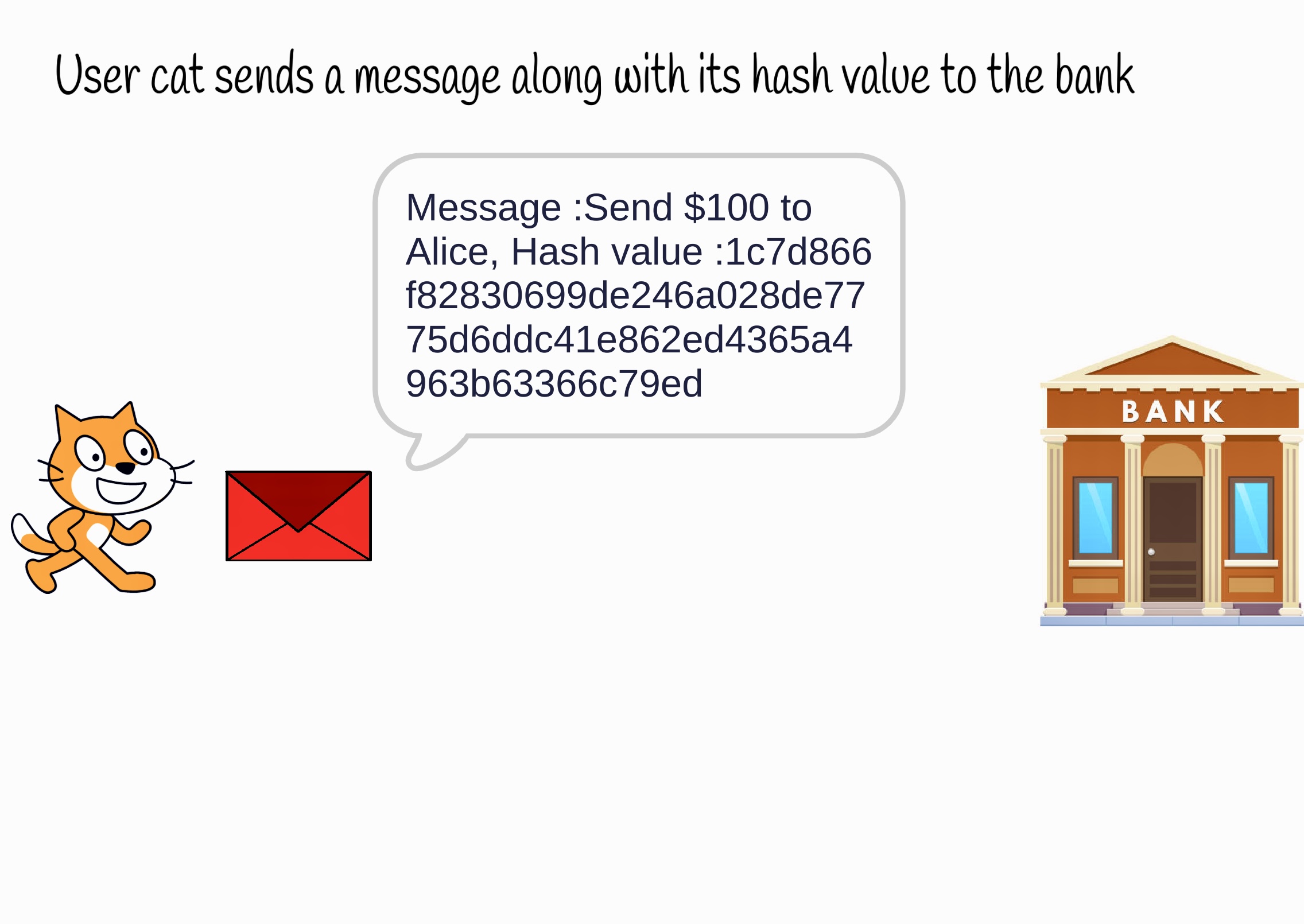}
        \caption{User cat sends a transaction message and associated hash value to a bank.}
        \label{fig:scenario2a}
      \end{subfigure}
     \hfill
     \newline
     \newline
     \begin{subfigure}[b]{0.43\textwidth}
        \centering
        \includegraphics[width=0.85\linewidth]{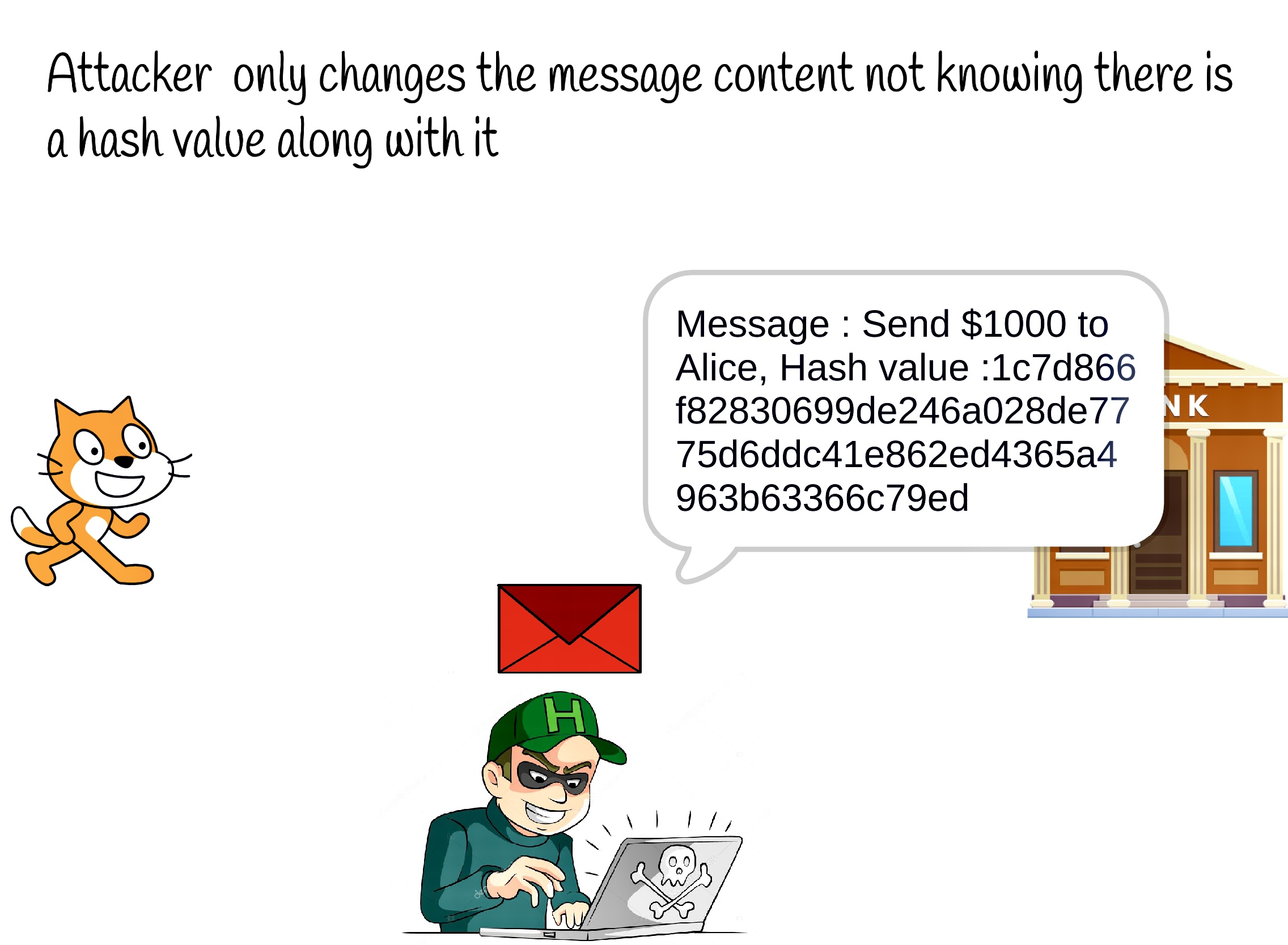}
        \caption{Attacker only modifies the transaction message not changing the associated hash value.}
        \label{fig:scenario2b}
      \end{subfigure}
     \hfill
    \newline
    \newline
    \begin{subfigure}[b]{0.43\textwidth}
        \centering
        \includegraphics[width=0.85\linewidth]{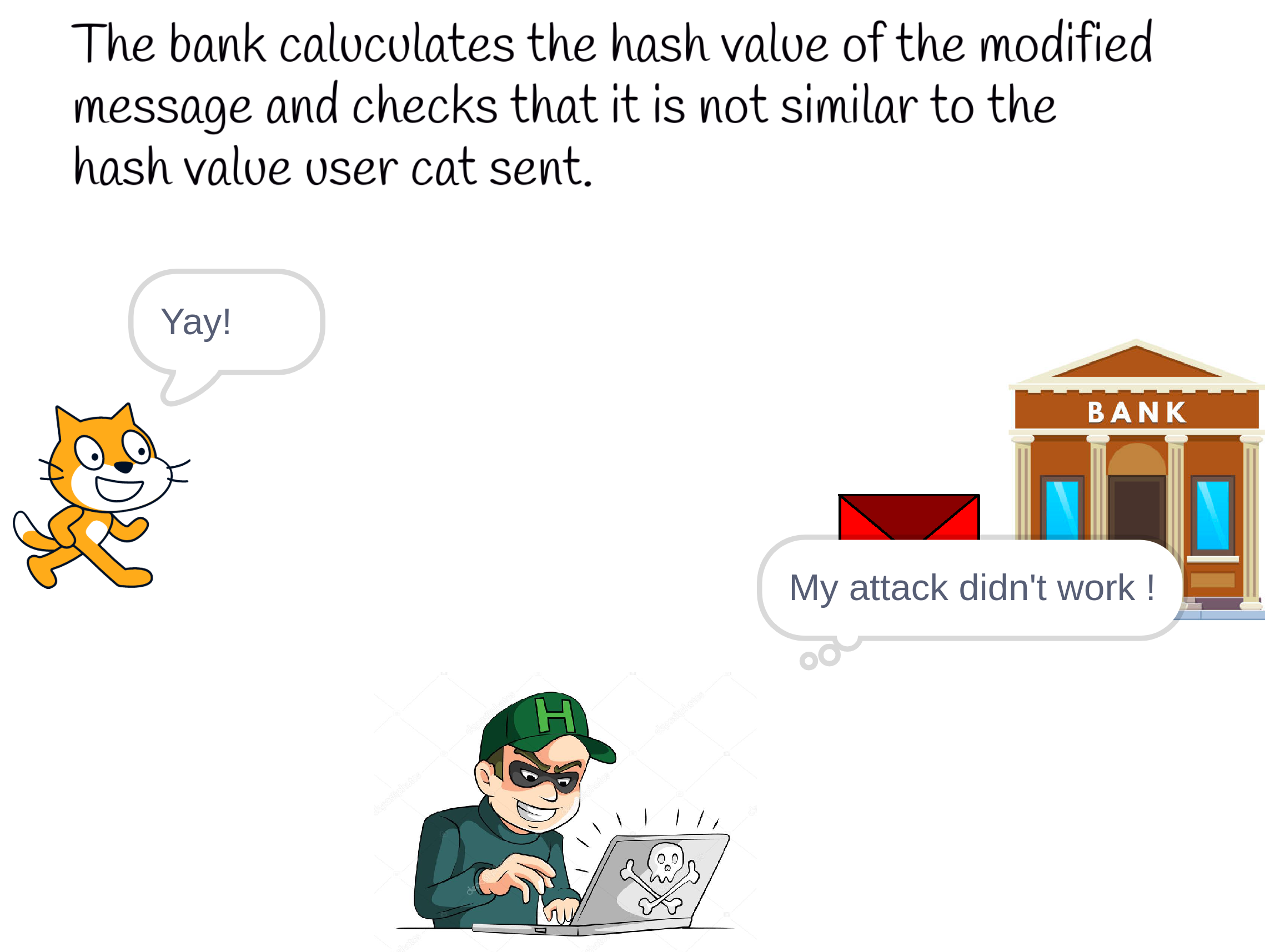}
        \caption{Bank recalculates the hash value of the message and matches it with the unmodified hash value.}
        \label{fig:scenario2c}
    \end{subfigure}
      
    \caption{A scenario that demonstrates Integrity Protection. Here, a user (cat) sends a transaction and associated hash value to the bank. An attacker changes the transaction but ignores the hash value. The attack fails as the bank recognizes the change owing to the hash value and declines the transaction.}
    \label{fig:Scenario2}
\end{figure}

We designed our curriculum using a Scratch extension called CryptoScratch \cite{cryptoscratch} that implements modern cryptographic algorithms like AES and RSA for encryption/decryption and the SHA-256 algorithm for message hashing. The CryptoScratch extension implements these algorithms as visual blocks with the same capabilities as conventional Scratch blocks. In addition, since Scratch has the capacity for visual demonstrations and a user interface that allows children to interact easily with the blocks, we determined that it would be a suitable platform for developing cybersecurity curricula. We note that the authors of CryptoScratch mostly focused on tool development and did not present any learning materials for the users of their system. Our focus, on the other hand, is on teaching how to use the aforementioned algorithmic blocks to provide security services such as confidentiality, integrity protection, and authentication. %To that end, 
%designing a cybersecurity curriculum utilizing the visual features of the Scratch environment and the functionality of cryptography blocks. As a pivot contribution, we want to prioritize on key security principles like as confidentiality, integrity, authentication, and anti-replay.
%we reduced the number of cryptographic blocks from the CryptoScratch system to make them more straightforward and easily understood. 

Using the CryptoScratch blocks, we designed scenarios and visual demonstrations that simplify the aforementioned core cybersecurity concepts.
%We applied these key cybersecurity concepts to a real-world scenario and developed a self-expressive system that demonstrates visually how these concepts are utilized to prevent/mitigate assaults. 
Our scenarios consider a conversation between a user and a bank, and an attacker attempting to intercept, modify and/or fabricate the conversations. Through collaboration between a diverse team comprising high-school students, undergraduate students, and security and education researchers, we designed a total of six scenarios that demonstrate the story of an attacker attacking bank transactions while the user and bank attempt to defend against these attacks using the aforementioned core cybersecurity concepts. The scenarios have been designed such that they start with a bank transaction that has no security protections (Scenario 1), then integrity protected using secure hashing algorithms (Scenario 2), then made confidential using symmetric cryptography (Scenario 4), and then authenticated using Certificate Authorities (Scenario 6). The other scenarios (Scenarios 3 and 5) demonstrate attacks that necessitate the need for the protections added in the other scenarios (Scenarios 2, 4, and 6). All these scenarios are discussed below. We note that our curriculum currently concentrates on teaching fundamental principles for the sake of simplicity for young children. According to national cyber security standards for K-12 students, we leave next advanced ideas such as replay attacks, reflection attacks, Man-in-the-Middle (MitM) attacks, Public Key Infrastructure (PKI) to explain public key cryptography, modern schemes, Message Authentication Codes, Authenticated Encryption, and the design of protocols such as Transport Layer Security (TLS) for future work.

\vspace{0.2em} \noindent \textbf{\textit{Notations}} - To explain our scenarios, we will denote a bank transaction message as $M$ throughout this section. Our goal is to demonstrate to students how to protect $M$ using the concepts of confidentiality, integrity protection, and authentication. The message $H(M)$ will denote a secure hashing algorithm such as SHA-256 on message $M$. The message $C=E(M, K)$ will denote encryption of message $M$ using a symmetric cryptographic algorithm such as AES and symmetric key $K$, and $\hat{M}=D(C, K)$ will denote decryption of ciphertext $C$ using the symmetric cryptographic algorithm and key $K$, such that $\hat{M}=M$.

\vspace{0.2em} \noindent \textbf{\textit{Integrity Protection}} - The first two scenarios of our story-based curriculum focus on teaching the concept of integrity protection. Integrity protection is one of the core cybersecurity principles, as it assures that data has not been altered during transmission between a sender and recipient. 
%The first two scenarios illustrate the possibility of an attack and the significance of data integrity in preventing it. 
In \textit{Scenario 1}, a user sends the sensitive transaction $M$ to the bank over an insecure channel. The attacker, intercepting this channel, modifies the transaction  $M$ to $M^{’}$ and forwards it to the bank. The bank receives the modified message $M^{’}$ and processes it resulting in a breach of the integrity of the transaction (see Fig.~\ref{fig:Scenario1}). The goal of this scenario is to show the dangers of insecure communications in critical everyday applications such as banks. In continuation to the previous scenario, Fig. \ref{fig:Scenario2} demonstrates \textit{Scenario 2} where the user and the bank recognize that the attacker has altered the message and discuss the use of message hashing as a defense strategy. To ensure the integrity of a future transaction, this time, the user transmits the message $M, H(M)$ to the bank. An attacker, unaware of the addition of message hashing, modifies the message $M$ to $M^{’}$ but leaves the hash as $H(M)$ to send $M^{’}, H(M)$ to the bank. The bank receives the modified $M^{’}$ and hash $H(M)$, verifies that the hash  $H(M^{'})$ is not the same as $H(M)$, and determines that the message has been altered (see Fig.~\ref{fig:scenario2c}). We note that this scheme is currently susceptible to an attacker changing the message and recalculating the hash. This vulnerability motivates the addition of the concept of confidentiality in the next scenarios.

\begin{figure}[t]
    \begin{subfigure}[b]{0.43\textwidth}
        \centering
        \includegraphics[width=0.8\linewidth]{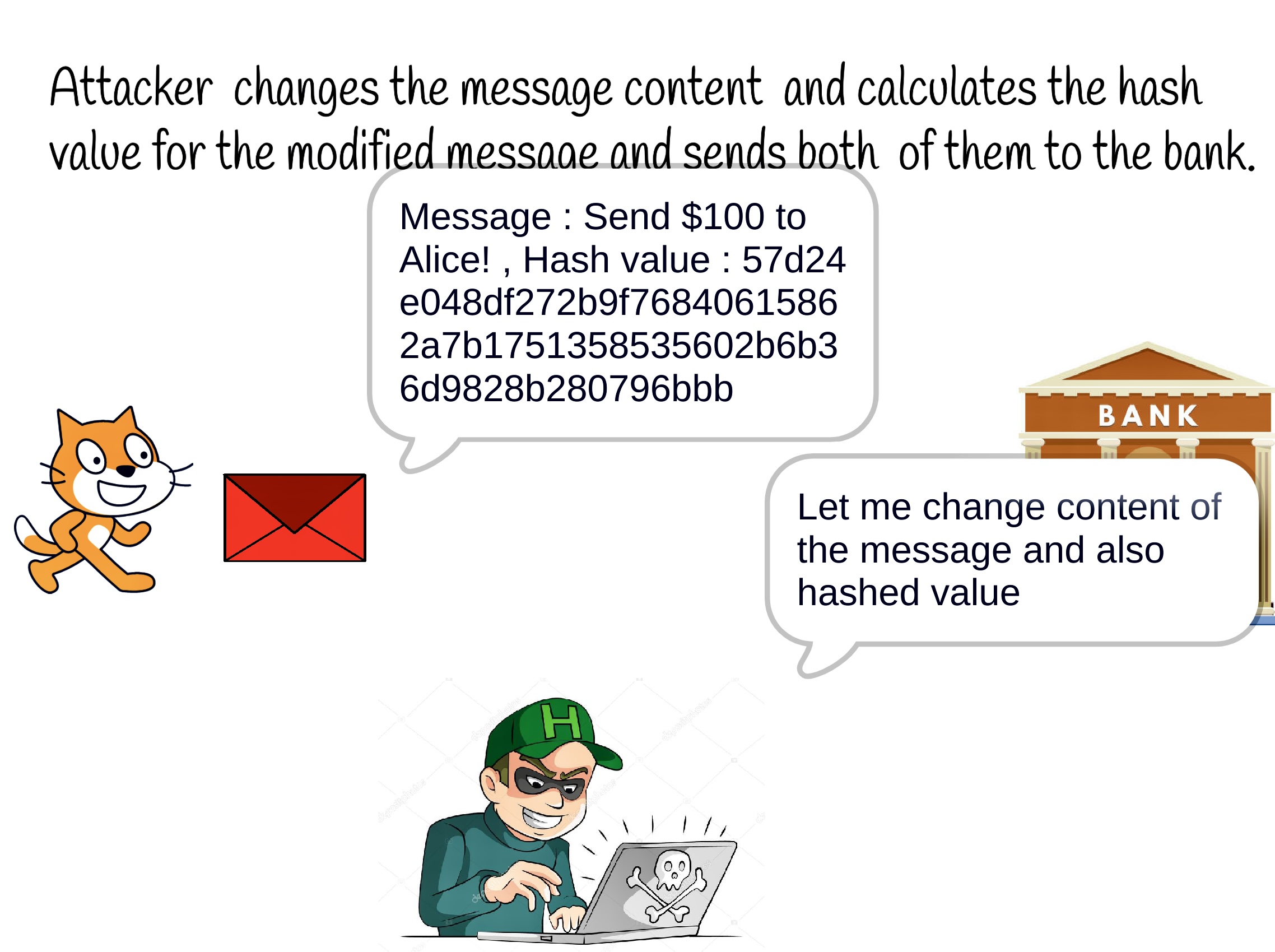}

    \caption{A attacker understands that user cat and bank are using message integrity, as new attack strategy he modifies the message and hash value}
    \label{fig:Scenario3a}
      \end{subfigure}
     \hfill
    \newline
     \begin{subfigure}[b]{0.43\textwidth}
        \centering
        \includegraphics[width=0.8\linewidth]{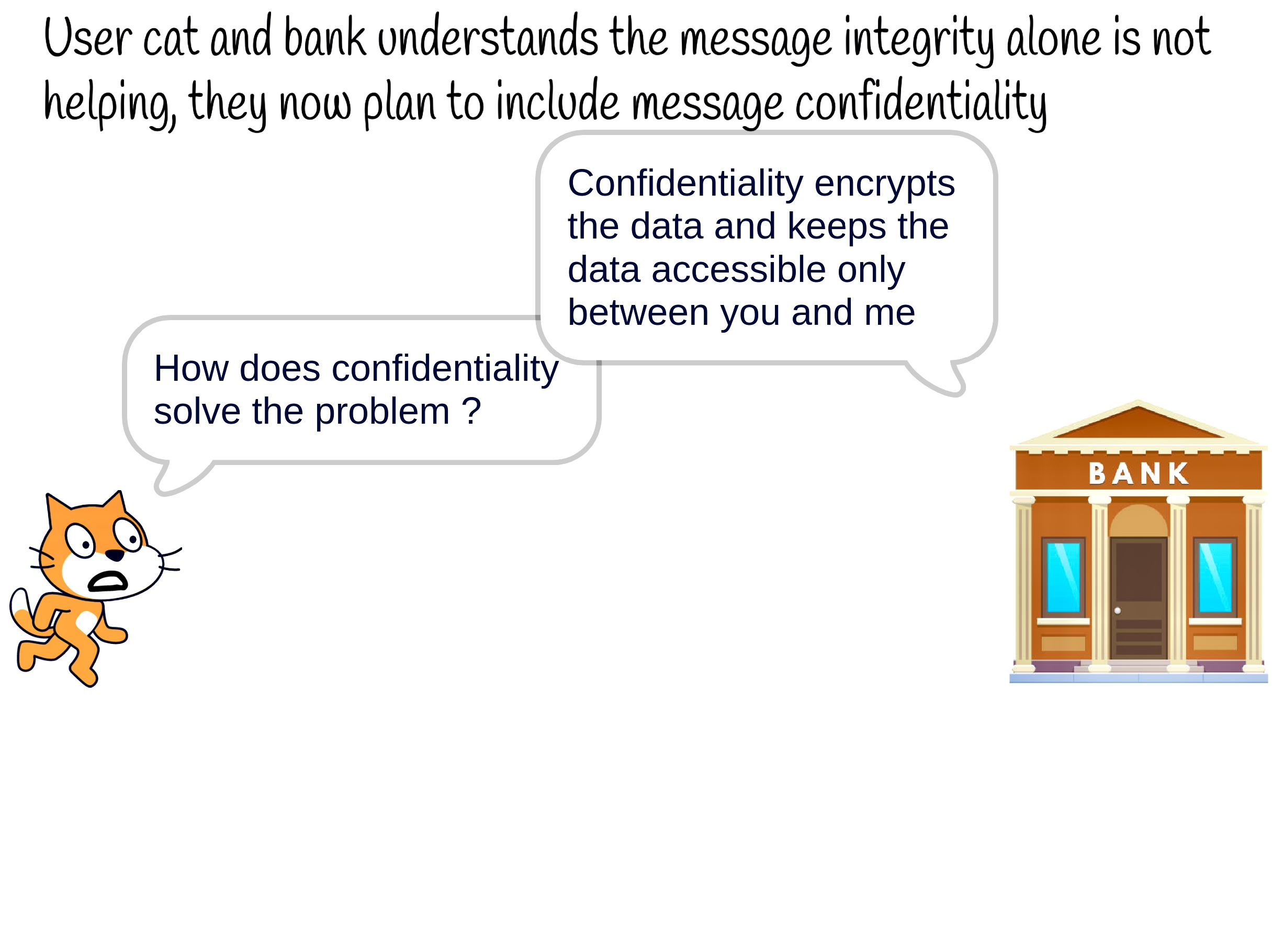}

    \caption{The user cat and bank include confidentiality, to mitigate modification of the messages .}
    \label{fig:Scenario3b}
      \end{subfigure}
    
     \hfill
     \newline
  
     \begin{subfigure}[b]{0.43\textwidth}
        \centering
        \includegraphics[width=0.8\linewidth]{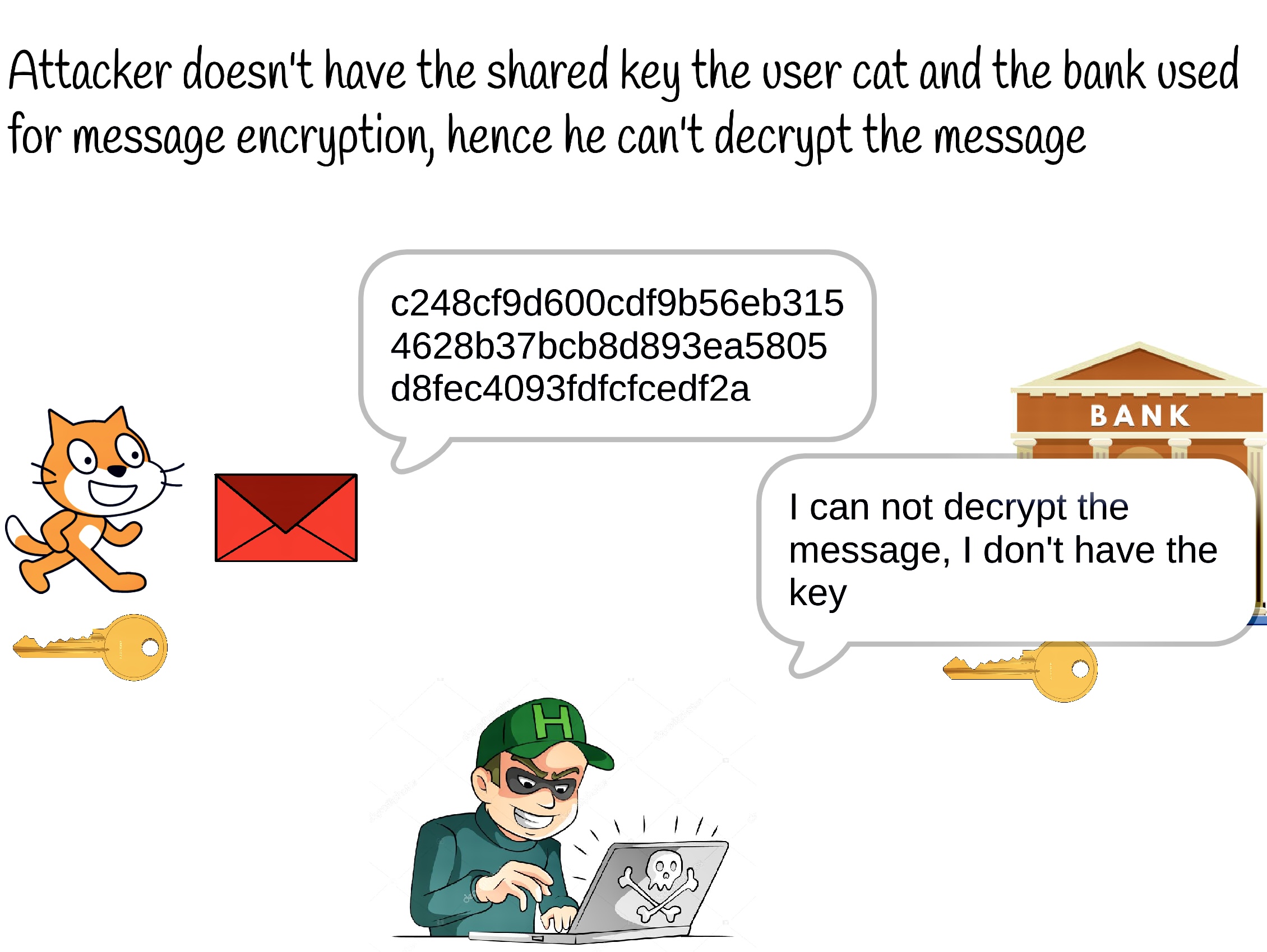}

    \caption{ The attacker can not decrypt the message and change the contents since he doesn't have the key.}
    \label{fig:Scenario3c}
      \end{subfigure}
     \hfill

    \caption{A scenario that demonstrates confidentiality. The attacker breaches message integrity protection by changing the user's message and the hash value sent to the bank. In order to mitigate this, the user cat and bank integrate message confidentiality and encrypt the message using the shared key.}
    \label{fig:blockcodes}
\end{figure}

 \textbf{\textit{Confidentiality}} - Scenarios 3 and 4 of our curriculum focus on confidentiality, which is a key cybersecurity concept that describes the ability of senders and recipients to protect information from unauthorized entities. We note that, even though confidentiality can be achieved using symmetric and asymmetric cryptography, we focus on symmetric cryptography for this work. This is because symmetric cryptography is more intuitive for K-12 students, while asymmetric cryptography can be slightly more advanced. We leave the exploration of simplifying asymmetric cryptography for students for a future extension of this work.
%These scenarios highlight the vulnerability that existed in the previous examples, with an emphasis on confidentiality as a solution. 
Continuing from Scenario 2, in \textit{Scenario 3}, the attacker discovers that the user and bank are now using message hashing to detect modifications to the message $M$.
%finds the communication technique used by the user and the bank  
This time, when the user sends the bank a message $M,H(M)$, the attacker attempts to disrupt communication by altering the hash value $H(M^{'})$ and sends $M^{’},H(M^{’})$ to the bank. The bank receives the new message, compares the hash, and accepts the transaction as it has no means of verifying that the message $M^{’}$ has been modified by the attacker. This attack is meant to demonstrate to students that integrity protection by itself cannot protect sensitive transactions, and must be combined with other core concepts, such as confidentiality for improved security. \Cref{fig:Scenario3a} illustrates the attack. 
%The bank generates the hash for the received message $H(M^{'})$ and validates it with the received hash value $H(M^{'})$ .
%Since both are equal, the bank cannot determine if the message has been altered.
In \textit{Scenario 4}, the user and the bank again discuss the attack and understand that integrity protection alone cannot secure the transaction as shown in Fig.~\ref{fig:Scenario3b}. To protect their future transaction from the attacker, they decide to use symmetric cryptography and share a secret key $K$. Let $\mathcal{T} = (M,H(M))$ denote the transaction message from Scenario 2. To protect the attack on $\mathcal{T}$, this time the user encrypts the message $\mathcal{T}$ to produce a ciphertext $C=E(\mathcal{T}, K)$ and transmits $C$ to the bank. The bank receives the ciphertext $C$ and decrypts it to get back the original transaction $\hat{\mathcal{T}} = D(C, K)$ such that $\hat{\mathcal{T}} = \mathcal{T}$.
%$E(M)$ and its hash value $E(H(M))$ using  the key $K$  . The bank decrypts the message $ D(E(M))$ and its hash value $D(E(H(M)))$ and verifies if the message has been updated.It generates hash of decrypted message $H(D(E(M))$ and verifies with hash value that was received $D(E(H(M)))$ are identical. 
Since the attacker does not have key $K$ to encrypt or decrypt the message $\mathcal{T}$, they cannot read this transaction preserving the confidentiality as shown in Fig.\ref{fig:Scenario3c}. Since $\hat{\mathcal{T}} = \mathcal{T}$, the bank also knows that the integrity of the message is preserved. However, we note that this communication is still vulnerable because it does not contain any means of verifying the user. The next scenario explains the potential risk of not validating the user and how authentication can be used to mitigate such risks.

\begin{figure}[t]
  \begin{subfigure}[b]{0.43\textwidth}
        \centering
        \includegraphics[width=0.8\linewidth]{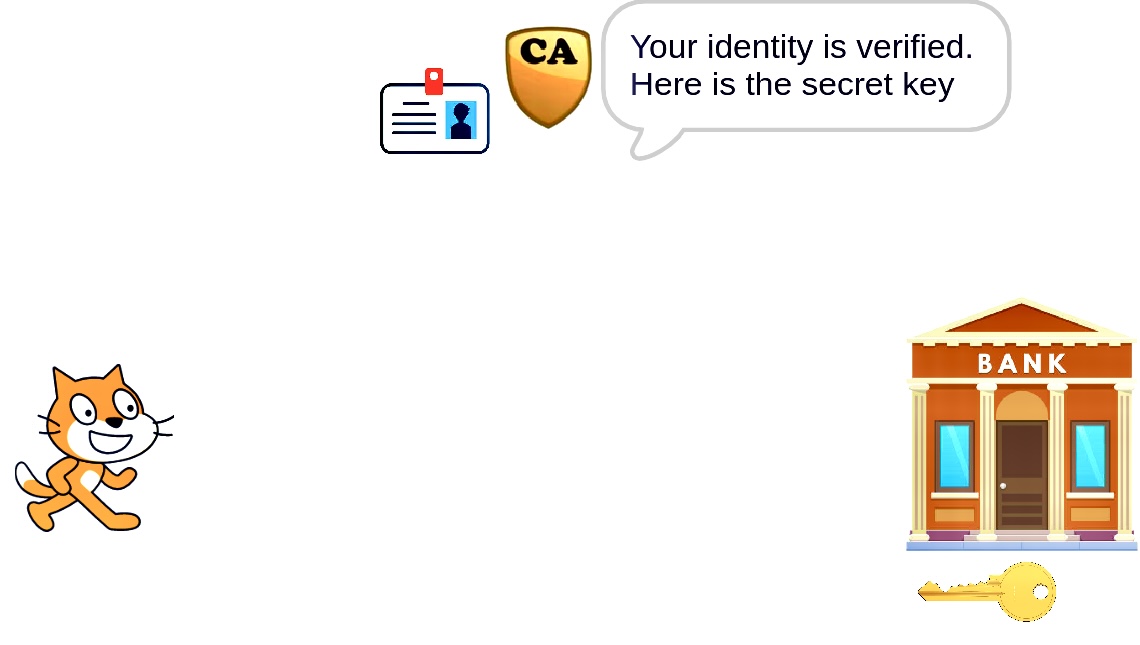}

    \caption{The Certificate Authority verifies the user identity and shares a secret key }
    \label{fig:auth1}
      \end{subfigure}
     \hfill
    \newline
     \begin{subfigure}[b]{0.43\textwidth}
        \centering
        \includegraphics[width=0.8\linewidth]{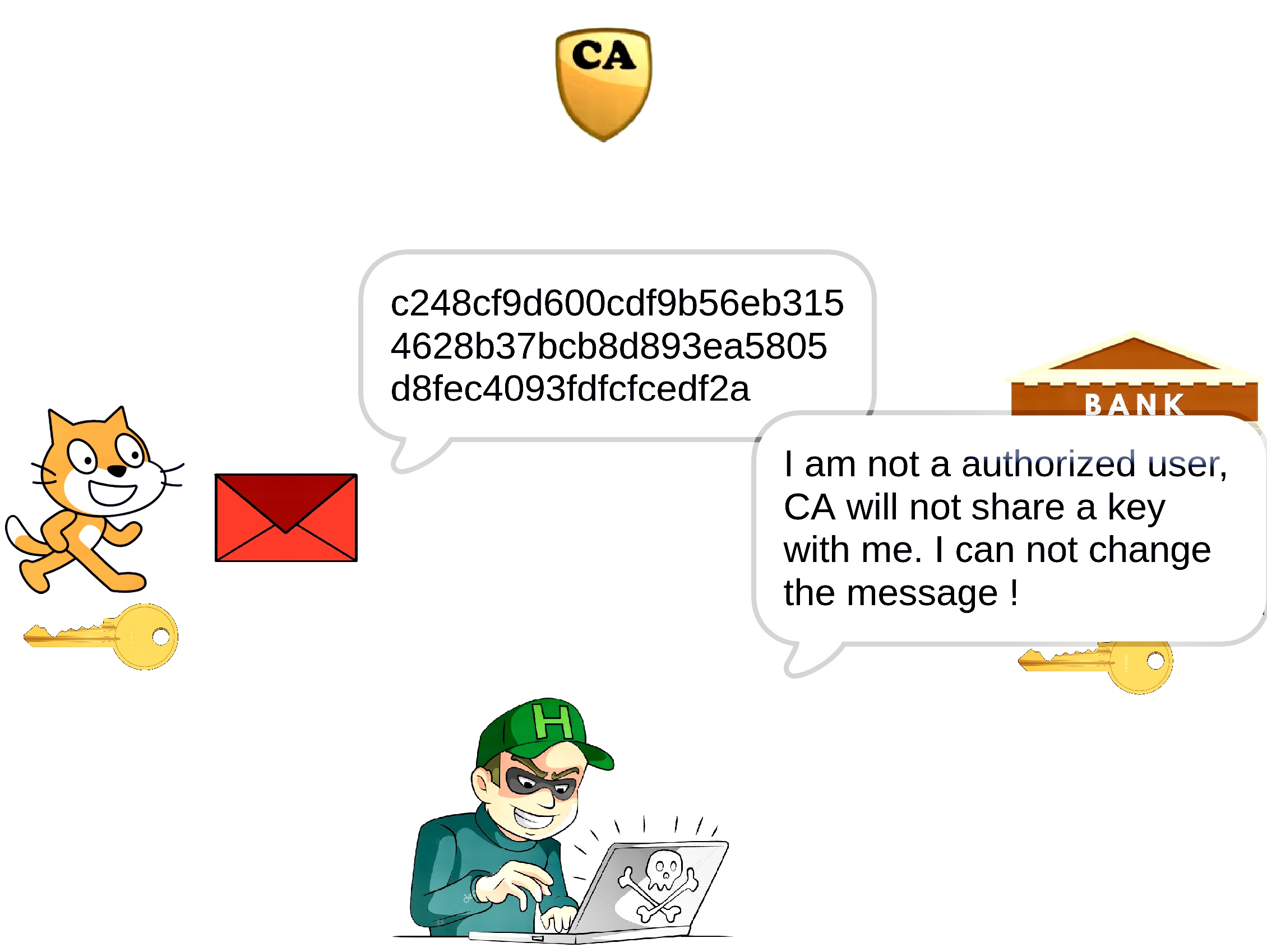}

    \caption{The user cat and bank use the secret key to encrypt and decrypt. However, an attacker can not access the message}
    \label{fig:auth2}
      \end{subfigure}
  \caption{A Certificate Authority (CA) verifies the identity of a user before a secret transaction key is sent to them via  a secure channel. The user cat and bank uses that secret key to encrypt and decrypt the message.}
  \label{fig:authentication}
\end{figure}

\vspace{0.2em} \noindent \textbf{\textit{Authentication}}: Our next two scenarios (Scenarios 5 and 6) focus on demonstrating the importance of learning authentication in a cybersecurity curriculum. Authentication is the process of identifying and validating the identity of a user and is the primary means used to control access to sensitive information (e.g., emails, and bank accounts) in the modern world. \textit{Scenario 5} demonstrates that an attacker can fabricate transactions even with the introduction of symmetric cryptography in Scenario 4. The attacker does this by impersonating the authorized user and obtaining the secret key $K$ from the bank. %Here, the bank share the key $K$ without validating the identity of the other participant. 
Now, the attacker simply needs to fabricate a new transaction $M^{'}$, encrypt the transaction $C^{'} = E((M^{'},H(M^{'})), K)$, and then transmit $C^{'}$ to the bank to perform a transaction as the user.
%Now, when a user sends an encrypted message $E(M)$ and $E(H(M))$, since the attacker has the key $K$ he modifies and encrypts the modified messages  $E(M^{'})$ and $E(H(M^{'}))$.   Assuming that only the authorized user holds the key, the bank decrypts the updated messages $ D(E(M^{'})) $ and $D(E(H(M^{'}))) $ and initiates a transaction.
This vulnerability can be addressed using authentication where the user and bank verify each other's identity before performing the transaction. \textit{Scenario 6}  introduces the concept of Certificate Authority (CA) that can be used to establish trust in the system. We note that the concept of CA in the real-world has its base in public key cryptography. However, since our focus is on using the more intuitive symmetric cryptography, we have deviated from the actual usage of a CA in protocols such as TLS. In our scenario, we use the CA as a trusted agent that verifies the user and the bank and then assigns a symmetric key $K_{UB}$ to the user and the bank (as opposed to a certificate in the real world). The CA shares this key via a secure channel after verifying the identity of the user and the bank as shown in Fig.\ref{fig:auth1}. We note that $K_{UB}$ is a secret key specific to the user and the bank. Considering the transaction $\mathcal{T} = (M,H(M))$ from Scenario 4, the user now transmits $C=E(\mathcal{T}, K_{UB})$ to the bank, and the bank decrypts $C$ to get back the original transaction $\hat{\mathcal{T}}=D(C, K_{UB})$ such that $\hat{\mathcal{T}} = \mathcal{T}$.
%CA shares the key $V(K)$  with the user and bank on verifying their identities. 
Even when the attacker attempts to impersonate a user, it fails in the verification process.
%A user encrypts the message E(M) and hash value of the message $E(H(M))$ with the verified key $V(K)$ . 
As such, the attacker cannot read, modify and/or fabricate the transaction message resulting in a scheme that provides all three security services - confidentiality, integrity protection, and authentication. Fig.\ref{fig:auth2} demonstrates this scenario where an attacker can not change the message contents.

\begin{figure}[t]
  \centering
  \includegraphics[width=0.47\columnwidth]{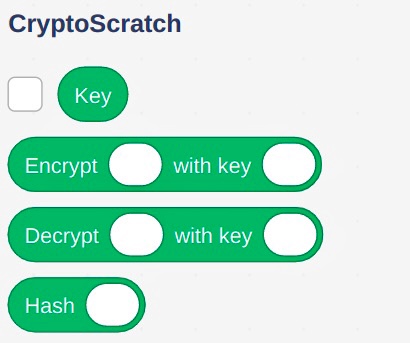}
  \caption{Cryptography blocks utilized to build the scenarios.}
  \label{fig:crypto}
\end{figure}

\vspace{-0.2em}
% Figure \ref{fig:blockcode} illustrates how these scenarios are implemented in the backend to present the visual curriculum to the students. 
\vspace{0.2em} \noindent \textbf{\textit{Implementation Details}}:
We explain the implementation of the curriculum by using the message integrity scenario (Scenario 2). The authors of CryptoScratch \cite{cryptoscratch} use the algorithm names such as AES and SHA for their block names. As we are building the curriculum for K-12 students, we have simplified the blocks to just refer to the key function of the block as shown in Fig. \ref{fig:crypto}.

\begin{figure}[t]
 
    \begin{subfigure}[b]{0.43\textwidth}
        \centering
        \includegraphics[width=0.78\linewidth]{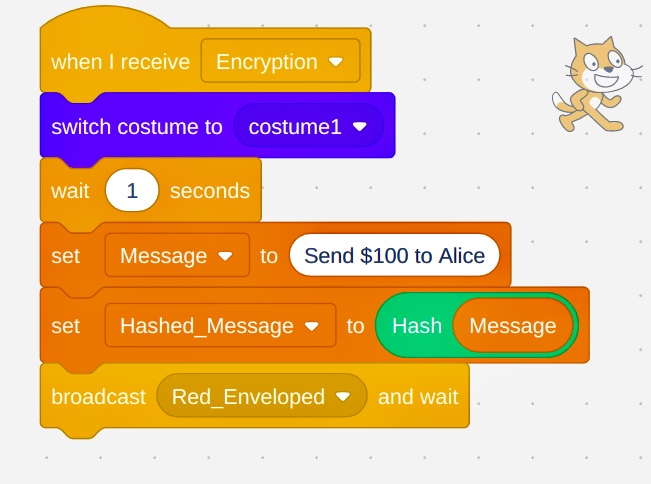}
        \caption{ Block code associated with a sprite user cat in the message integrity scenario}
        \label{fig:blockcode1}
      \end{subfigure}
     \hfill
    \newline
     \begin{subfigure}[b]{0.43\textwidth}
        \centering
        \includegraphics[width=0.9\linewidth]{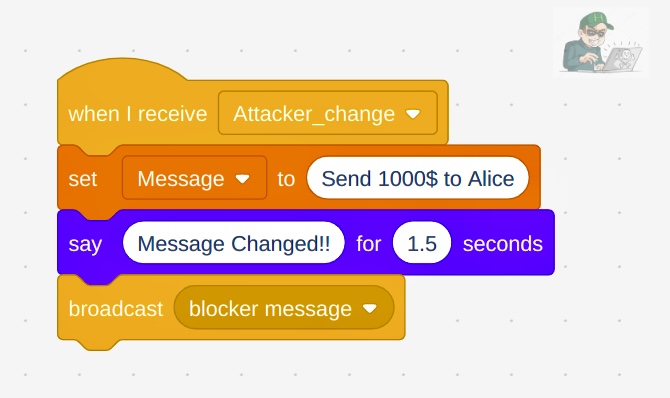}
        \caption{Block code associated with a sprite attacker in the message integrity scenario}
        \label{fig:blockcode2}
      \end{subfigure}
     \hfill
     \newline
     \newline
     \begin{subfigure}[b]{0.43\textwidth}
        \centering
        \includegraphics[width=0.9\linewidth]{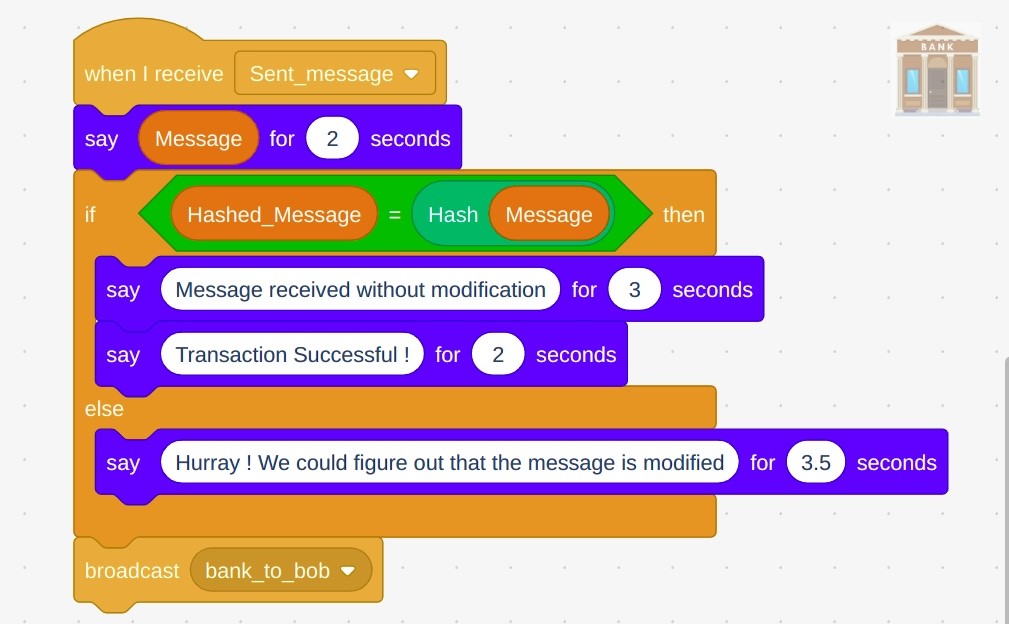}
        \caption{Block code associated with a sprite bank in a message integrity scenario} 
        \label{fig:blockcode3}
      \end{subfigure}
     \hfill
    \caption{Executable block code for sprites involved in the message integrity scenario }
    \label{fig:blockcodes}
\end{figure}

We begin the flow of the scenario with user cat, where the user initiates the transaction and executes the associated sprite code as shown in Fig.\ref{fig:blockcode1}. This sprite code first creates a transaction message and saves it in a Scratch variable called `Message'. Then it computes the hash of the message and saves it in a Scratch variable called `Hashed\_Message'. Once the message and the associated hash value are generated, the user transmits the message using the broadcast block. The broadcast block is similar to a function call in conventional programming. As a continuation, the attacker sprite executes its associated sprite code, this code modifies only the transaction message which is modifying a variable `Message' in scratch, unknowing there is an associated hash value as shown in Fig. \ref{fig:blockcode2}. Now, the adversary re-transmits the information to the bank using a broadcast block, which in turn calls/triggers the code associated with the sprite bank. The associated code recalculates the hash value for the modified message and compares it to the unmodified hash value using the comparison operator block as shown in Fig. \ref{fig:blockcode3}. Furthermore, using a scratch logic block on these comparisons assesses if the transaction is legal and determines whether to discard or execute the transaction.  Note that here `Red\_enveloped', and `Blocker\_message'  that are used in broadcast blocks as shown in Fig.\ref{fig:blockcodes} are temporary block code to create flow in visualization, and does not implement logic, due to space constraint we didn't include those blockcode in the images.

All the aforementioned principles were explained by relating them to real-world scenarios and making them highly visual through the conversation between the user, bank, and the attacker. We believe that these qualities of our curriculum aid in the retention of these concepts in children's minds. A video demonstration of these scenarios is also available \footnote{\url{https://www.dropbox.com/s/cq4031sx3ore50t/Video.mp4?dl=0}\label{video}}.
%We believe that a visual presentation improves kid's level of understanding and cognition. 
In order to assess the curriculum, we conducted a user study, and the next section is dedicated to analyzing the study's findings.

\section{\textbf{Evaluation}}
We conducted a user study to assess  our curriculum competency in teaching students the cybersecurity concepts of confidentiality, integrity protection, and authentication. 
%We dedicated this section to present our analytical findings from the evaluation study with eighteen K-12 students. 
This section presents the user study design, recruitment of participants, and results demonstrating the student's perception of our curriculum. 
%Further discussions are structured as below.

\subsection{ \textbf{Design of the Evaluation Study}}
The user study’s aims are threefold: 
\begin{itemize}
  \item  To assess the accessibility of the curriculum
  \item To assess the effectiveness of the curriculum
  \item  To assess the impact on students’ learning
%. In achieving our aims, we hope to contribute to teaching the cybersecurity curriculum to K-12 students.
\end{itemize}
\vspace{0.3em} \noindent
\subsubsection{\textbf{Recruitment}}Study participants were recruited via a snowball sampling method ~\cite{snowball} , frequently used in recruiting participants in various disciplines. One of the authors created a flyer that advertised the study and sent it to acquaintances, friends, and former colleagues after securing approval from the authors' university's Institutional Review Board (IRB). As a result of the recruitment effort, we recruited middle-school and high-school students. While a few middle-school students were directly enrolled in the study, most students were enrolled by their parents. Because the study participants are minors, we received a parental consent form from every participant, which is required by the authors’ university’s IRB. The first author contacted the students/parents via email and confirmed their placement for the study. Then, the author shared the workshop details with the study participants.
\vspace{0.3em} \noindent
\subsubsection{\textbf{Study Design}} The participants were part of a 3-hour one-day virtual workshop, which entailed lectures and hands-on training sessions on the core cybersecurity concepts described in Section~\ref{sec:technical}.
Due to the continuing COVID-19 pandemic, most of the parents preferred to have the workshop conducted remotely ~\cite{useful}. As such, we held the workshop virtually. The workshop was conducted synchronously to mitigate any difficulties that may have arisen from the virtual teaching/learning environment. 
%Synchronous learning can create an enjoyable real-time interaction with instructors and peers (Mahalakshmi and Radha, 2020).
The workshop began with a brief introductory session to create a friendly atmosphere and promote participant engagement. Following everyone's introductions and discussion of their interests, we requested the participants to complete our pre-survey questionnaire. The pre-survey questions include their socio-demographic questions, prior knowledge of computer programming skills, and cybersecurity concepts.

The first hour of the workshop was dedicated to educating about cyber awareness and cybersecurity basics. Some of the most prevalent attacks, such as MitM, phishing, password, and malware were explained in detail. We also prioritized the teaching of cybersecurity basics such as securing passwords and avoiding spam emails to avoid cybercrimes. The rest of the workshop focused on teaching confidentiality, integrity protection, and authentication from our curriculum. The students were shown the demonstrations of those scenarios, followed by discussions of what the next potential attack can be and their ideas about how those attacks can be defended. Then, the participants were asked to solve a technical survey comprising ten questions assessing their understanding of the concepts.

\vspace{0.15em} \noindent
\subsection{\textbf{Pre- and Post-Survey Evaluation}}
Our goal with the pre and post-survey was to understand the curriculum’s accessibility and contribution to the students’ learning about cybersecurity. We wanted to assess whether students enjoyed using the interface to learn cybersecurity. In doing so, we wanted to provide students an opportunity to be prepared and aware of cybersecurity. A strong understanding of core knowledge is fundamental to students for learning and also working on complex cybersecurity issues in the future.
\vspace{0.3em} \noindent
\subsubsection{\textbf{Participant Demographics}}

In total, we had 18 study participants where 10 (55.5\%) are female and 8 (44.5\%) are male.
All our study participants were aged between 10 to 15 years old (see Table~\ref{tab:study_dist} for distribution). 
 %Around 22.3\% participants were 12 years old, while more than half of the participants were 11, 14, and 15 years old. 
 All of them studied in the fifth to eleventh grades. Our participants were predominantly Asian (94.45\%) with 1 Caucasian student. Almost all of them indicated that they were new to cybersecurity.
 %The remaining percentage  is Caucasian . Table 1 presents age group  statistics of the 18 participants.
%We believed it was important to provide a positive learning environment and to encourage female participants to express themselves throughout the workshop. We are pleased that women are encouraged and enticed to participate in cybersecurity field.

% \textbf{Table 1. Gender statistics}
% \newline
% \begin{center}
% \begin{tabularx}{0.2\textwidth} { 
%   | >{\centering\arraybackslash}X 
%   | >{\centering\arraybackslash}X 
%   | >{\centering\arraybackslash}X|}
%  \hline
%  Gender & N & \% \\
%  \hline
%  Male & 8 & 44.5 \\
%  \hline
%  Female & 10 & 55.5 \\
%  \hline
%  Total & 18 & 100.0 \\
%  \hline
% \end{tabularx}
% \newline
% \end{center}

\begin{table}[t]
    \small
    \caption{Participants age distribution.}
% \newline
    \centering
\begin{tabularx}{0.3\textwidth} { 
  | >{\centering\arraybackslash}X 
  | >{\centering\arraybackslash}X 
  | >{\centering\arraybackslash}X |}
 \hline
  \textbf{Age} & \textbf{N} & \textbf{\%} \\
 \hline
  10 & 1 & 5.56  \\
 \hline
  11 & 2 & 11.13  \\
 \hline
 12 & 4 & 22.2 \\
 \hline
 13 & 2 & 11.13 \\
 \hline
 14 & 5 & 27.78  \\
 \hline
 15 & 4 & 22.2 \\
 \hline
 Total & 18 & 100.0  \\
 \hline
\end{tabularx}\label{tab:study_dist}
\newline
\end{table}

Among the 18 participants, 10 (55.6\%) reported in the pre-survey that they were `interested', and 6 (33.3\%) of the participants as `very interested' to learn cybersecurity. 
%The remaining 11.1 percentage has been moderately interested towards learning cybersecurity. Fig. 6 shows the percentage of students who are interested to learn cybersecurity  
%  \begin{figure}[h]
%   \centering
%   \includegraphics[width= 7cm, height = 3.5 cm]{The_intrest.png}
%   \caption{The statistics of student's interest to learn cybersecurity}
%   \label{fig:interst}
% \end{figure}
In the post-survey questionnaire, every student unanimously voted that our cybersecurity curriculum is very useful and can explain complex cybersecurity concepts easily. Every student found this method of teaching cybersecurity principles to be very useful and effective.

\vspace{0.3em} \noindent 
\subsubsection{\textbf{Curriculum Assessment}}
\begin{figure}[t]
  \centering
  \includegraphics[width=0.84\columnwidth]{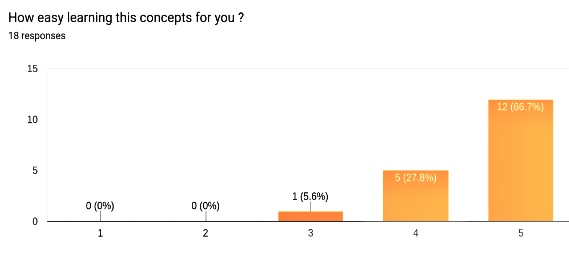}
  \caption{Percentage of students who found the curriculum easy to understand and learn the concepts.}
  \label{fig:easy}
\end{figure}

Our participants unanimously found the cybersecurity concepts from our curriculum easy to learn. Specifically, 12 (66.7\%) participants found the concepts very easy to learn (Likert scale 5), while 5 (27.8\%) participants found it quite easy to learn (Likert scale 4), as shown in Fig.~\ref{fig:easy}.

\begin{figure}[t]
  \centering
  \includegraphics[width=0.84\columnwidth]{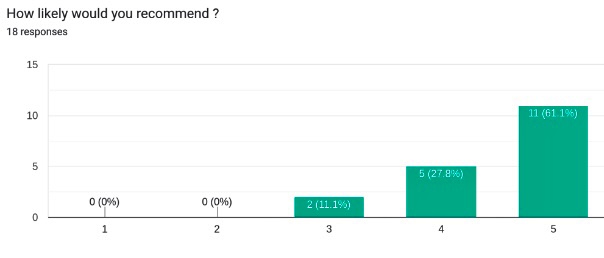}
  \caption{The likelihood of participants recommending this curriculum to other students.}
  \label{fig:recommend}
\end{figure}

Most of the students reported that they would recommend the curriculum to others. In summary, 11 (61.1\%) students responded that they would highly recommend the curriculum to other students, while 5 (33.3\%) students would recommend it to others. The remaining 2 students (11.1\%) indicated that they are happy to recommend it to others, as shown in Fig.\ref{fig:recommend}.

Overall, the students enjoyed interacting with our curriculum and left comments such as -
\textit{"I think I found it very useful as it provides a visual for me. I can see what is happening to the information in a very simple way which enables me to understand the concepts thoroughly and also apply them in the future."}, and \textit{"I enjoyed the live examples for the cybersecurity principles on Scratch!"}.

\vspace{0.3em} \noindent
\subsection{\textbf{Assessing Students' understanding}}
We asked the students to fill out a technical survey questionnaire to assess their understanding of cybersecurity concepts after learning such concepts in our curriculum. %To justify our second research contribution, 
In our workshop, we presented these concepts to the students for an hour and then asked them to answer the technical survey. This survey had ten multiple-choice questions, which were distributed over three sections. 
%All the questions worth 1 point if the students answered correctly. 
The results were promising. The average score of correct answers was 9.28 out of 10. We were also able to see that the female participants were eagerly and actively engaged in the workshop and discussions. It is also encouraging to see this result since gender gaps in cybersecurity education and practice still exist ~\cite{geneder}. Among the 10 female students, the mean of them answering technical questions correctly was 9.7 out of 10.  

\vspace{0.2em} \noindent
\textit{\textbf{Section 1}} - We had five conceptual questions in this section to evaluate the students' comprehension of the core concepts. The multiple-choice questions such as \textit{``Which of the following ensures that shared information is in a format that is not modified in transit?"} and 
\textit{``Which of following cyber security principle relates to encrypting the data?"} assessed basic understanding.
The percentage of accurate responses is shown in Table \ref{tab:table2}, allowing us to determine whether these scenarios adequately convey the meaning of each fundamental concept. The average score on these five questions was 4.33 out of 5.

\begin{table}[h]
    \small
    % \centering
    % \begin{tabular}{c|c}
    %      &  \\
    %      & 
    % \end{tabular}
    % \caption{Caption}

    \caption{Conceptual questions evaluation.}
% \newlineTable 2. Conceptual questions evaluation
    \centering
\begin{tabularx}{0.5\textwidth} { 
  | >{\centering\arraybackslash}X 
  | >{\centering\arraybackslash}X 
  | >{\centering\arraybackslash}X |}
 \hline
 \textbf{Topic} & \textbf{Correct} & \textbf{Incorrect}  \\
 \hline
 Confidentiality & 88.9\% (16 students) & 11.1\% (2 students)\\
 \hline
 Integrity & 94.4\% (17 students) & 5.6\% (1 student) \\
 \hline
 Authentication & 88.9\% (16 students) & 11.1\% (2 students) \\
 \hline
\end{tabularx}\label{tab:table2}
\newline
\end{table}

\begin{figure}[t]
  \centering
  \includegraphics[width=0.85\columnwidth]{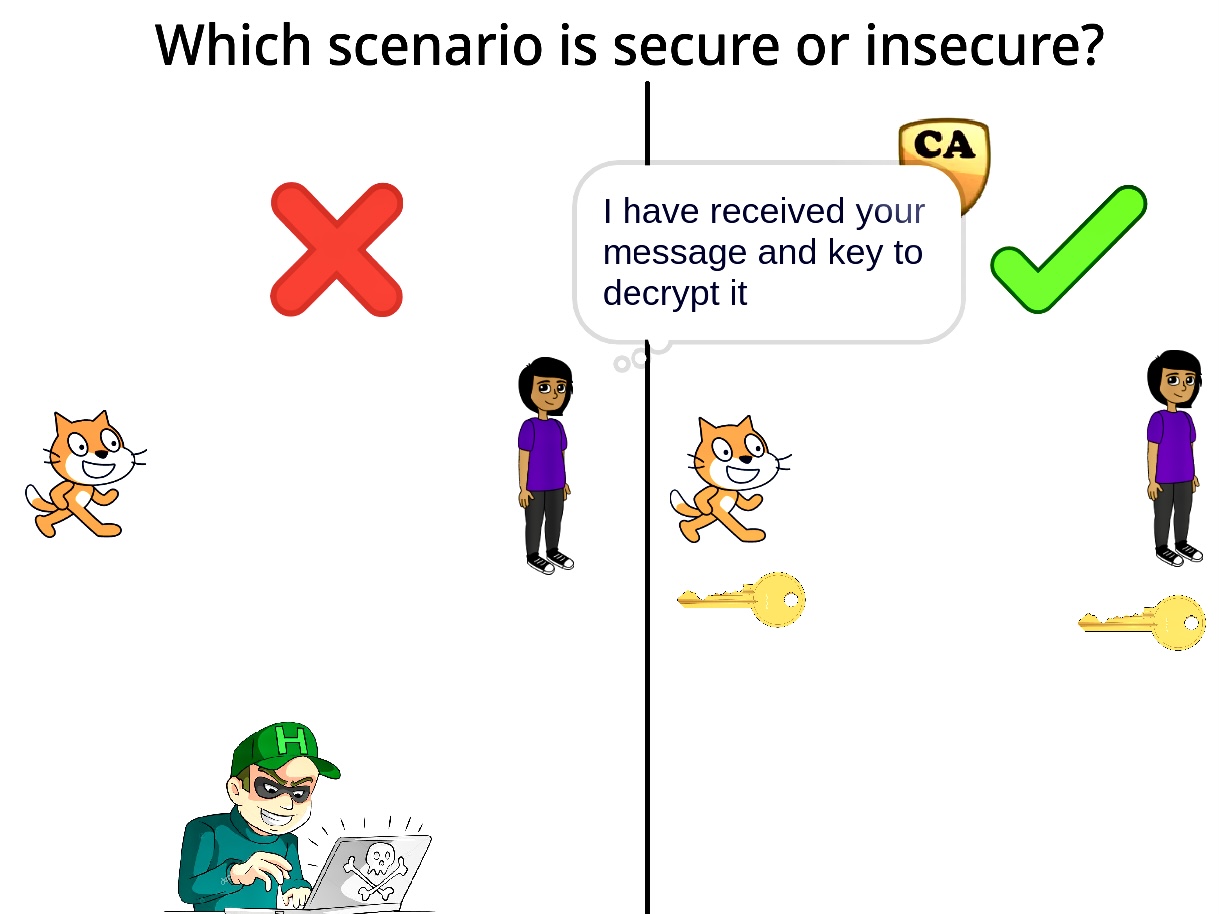}
  \caption{An example of a secure and insecure scenario shown to the students to assess their understanding of the curriculum.}
  \label{fig:secure}
\end{figure}

\vspace{0.2em} \noindent
\textit{\textbf{Section 2}} - We also showed the participants some secure and insecure visual scenario questions around cybersecurity concepts during the workshop. Fig.~\ref{fig:secure} demonstrates a simple visual scenario, on the right we show them a secure scenario, where CA shares the key with the authenticated parties on verifying their identities (i.e authentication and encryption). On the left, we show them an insecure scenario, where the user cat shares a message and also a key with it to the bank (i.e., no authentication). Note that students were already taught about concepts of authentication, encryption, CA key sharing in our curriculum. We believe that differentiating between these scenarios enables students to think about vulnerabilities in the environment. The mean of students answering these two questions correctly was 1.94 out of 2.

%We asked students to record their answers in survey forum, the students were engaged and lead healthy discussion among themselves to answer these questions. 

\begin{figure}[t]
   \centering
   \includegraphics[width=0.85\columnwidth]{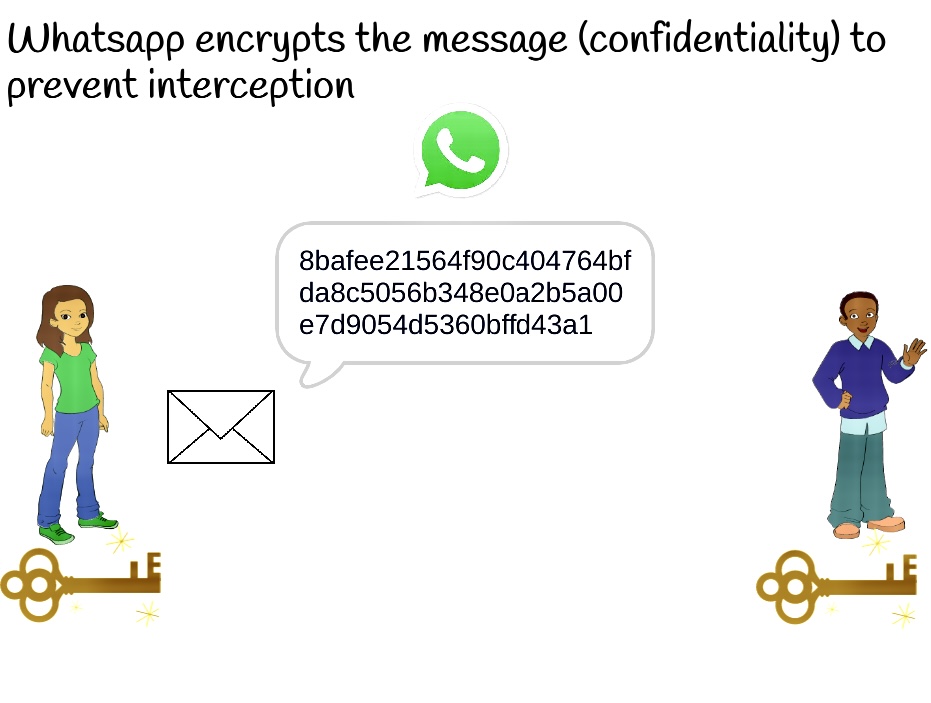}
   \caption{An example of a visual scenario shown to students to assess their understanding of the concepts translated into real-world contexts.}
   \label{fig:whatsapp}
\end{figure}

\vspace{0.2em} \noindent
\textit{\textbf{Section 3}} - This set of questions was included in the survey to see whether students were restricting their knowledge to the scope of the bank and user scenario they were taught or whether they understood the real-world implications of the learned concepts. To facilitate comprehension, these questions are framed in a manner that asks about the same concepts in many different contexts. Fig.\ref{fig:whatsapp} shows two individuals conversing on a messaging platform, and students were asked to name the fundamental concept that secures WhatsApp or message platform conversations. The students unanimously answered that WhatsApp employs confidentiality, convincing us that they could discern between the fundamental principles. The average value of students answering these questions is 2.94 out of 3. By these results, we believe that students can understand the concepts and relate the usage of these concepts in other real-world environments.

\section{\textbf{Potential Benefits}}

We believe that this real-world, visual, and narrative curriculum with a block-based programming language (i.e., Scratch) will help students understand the importance and existence of cybersecurity concepts to build secure systems. Our curriculum can intuitively address the question of why and how these concepts are used which is, in our opinion, essential for a thorough comprehension of the subject. Moreover, this may improve children's knowledge of cyberattacks and instill a sense of caution towards safeguarding their personal information in everyday life.
We anticipate that this will also help children grasp advanced cybersecurity principles at an early age (which is a subject of our future research), and help them build real-time secure infrastructure scenarios to learn cybersecurity on the Scratch platform. Students can thus master the application of cybersecurity concepts in the real world without dealing with complexities associated with programming languages or tools.

We believe that introducing our intuitive, user-friendly, and age-appropriate cybersecurity curriculum will significantly expand the uptake of cybersecurity education in K-12 institutions. The prevalence of Scratch in schools influenced our decision to develop our curriculum on the Scratch platform; consequently, integrating our curriculum into schools should be feasible for educators. We think that the free and multilingual availability of the  Scratch platform to anybody with an Internet connection will benefit students from diverse and underrepresented socioeconomic backgrounds.

\section{\textbf{Conclusion}}
This paper proposes an intuitive and real-scenario-based cybersecurity curriculum using Scratch to teach the core cybersecurity concepts of confidentiality, integrity protection, and authentication. The curriculum introduces students to three visual scenarios demonstrating attacks when systems do not integrate concepts such as confidentiality, integrity protection, and authentication. Then, it introduces them to three scenarios that build on the attacks to demonstrate how the fundamental concepts can be used to defend against them. Based on an evaluation survey, 67\% out of 18 middle and high-school students found the curriculum very easy, and 28\% found it relatively easy to learn and comprehend the concepts in our curriculum. The initial study shows the potential of our curriculum  and provides the impetus for the future to integrate other complex cybersecurity concepts into the Scratch platform.

%Our curriculum's ability to visually demonstrate attack and defensive scenarios will aid students in understanding such concepts and excelling in this field. We anticipate that in the future, if children were educated in more advanced cybersecurity principles at an early age, they would be able to build these real-time secure infrastructure scenarios in the Scratch platform. This enables students to comprehend the intuitive application of cybersecurity concepts in the real world without any programming or tool-specific complexities.

\bibliographystyle{IEEEtran}
\bibliography{refer}

\end{document}